# An analytic equation for single cell electrochemical impedance spectroscopy with a dependence on cell position

Yusuke Sugahara[1] and Shigeyasu Uno[1 a)]

[1]*Department of Electrical systems, Graduate School of Science and Engineering, Ritsumeikan University, Shiga 525-8577, Japan*

[a)]Author to whom correspondence should be addressed: suno@fc.ritsumei.ac.jp

## Abstract

An analytic equation for electrochemical impedance of a single-cell measured with a microelectrode is presented. A previously reported equation had a practical problem that it is valid only when the microelectrode resides at the center of the cell under test. In this work, we propose a new analytic equation incorporating dependence on cell position, and confirmed its effectiveness by numerical simulation. Comparisons show that our proposed equation gives an excellent agreement with simulated impedance values. Discrepancies between results from our equation and numerical simulation is suppressed within 13%, which is a dramatic reduction from the previously-reported equation as large as 58%. The proposed analytic equation is expected to enable more accurate analysis in actual cell experiments.

# I. Introduction

Advancing understanding in the medical and pharmaceutical sciences requires understanding biological phenomena and cell structure.[1,2] There is, therefore, a high demand for methods to monitor cultured cells. In recent years, optical techniques such as fluorescence microscopy and flow cytometry have been widely used to monitor cells.[3,4] For example, labeled substances in cells can be monitored via fluorescence microscopy,[5] which can be used to quantify cell viability as well as specific cellular components such as proteins.[5,6] An advantage of fluorescence microscopy is that it enables monitoring with spatial and temporal resolution.[7] However, use of this technique may adversely give negative impact to cells because of the binding of labeled substances to intracellular molecules.[8] In flow cytometry, a large number of labeled cells are passed through a tube, and individual cells are detected with a laser beam.[9] Flow cytometry can be used to measure various characteristics of individual cells, including their viability,[10,11] and the measurements can be made at a high throughput.[9,10] However, flow cytometry is not suitable for long-term monitoring because measured cells are not recultured. An individual cell can be monitored without labeling via the patch-clamp method, which involves measurement of ion channel biophysics and membrane properties by insertion of a glass electrode into a cell.[12,13] The patch-clamp method enables measurements of the characteristics of cell membranes and cell viability after drug administration.[14] The advantage of the patch-clamp method is that the individual characteristic of cell membrane can be obtained without labeling, but it potentially damages a cell, and the measurement throughput is extremely low.[15]

Electrochemical impedance spectroscopy (EIS) is another label-free method that can be used to monitor single cells. The EIS is a measurement to monitor the cell from the response signal when a small AC voltage is applied to an electrode.[16,17] Characteristics such as cellular permittivity, conductivity, cell size, and cell viability can be obtained by this method.[18-20] The advantage of EIS is that it enables analysis of live-cells in real-time under non-invasive, label-free conditions.[21-23] Conventional EIS measurements average cell characteristics among all cells in a population.[24] Because conventional EIS monitors the status of a population of cells instead of individual cell, it can ignore complicated interactions due to cellular heterogeneity.[25] Some characteristics such as genes expression and difference in the microenvironment of cells are difficult to measure.[26] Monitoring of single cells to capture differences between individual cell has therefore received attention.[27,28] A numerical simulation study reported that single-cell EIS enables clear observation of the conductivity and permittivity of a single cell when the electrode is smaller than the cell.[29] The smaller the electrode, the clearer the electrical characteristics will be because the flow of the electric current is concentrated on the single cell being examined. Such monitoring of single cells may help to clarify the causes of disease.[30] However, successful single-cell measurements require that cells be aligned on top of the microelectrode. The difficulty of aligning cells in this way can be overcome by using complementary metal-oxide-semiconductor (CMOS) technology to arrange a large array of microelectrodes on the sensor surface.[29,31-33] With such an array of electrodes, it is possible to find and choose an electrode under the cell of interest without controlling the position of the cell.

In the analysis of EIS measurements and optimization of sensor design, an analytic expression for impedance under specified conditions plays an important role. A well-known analytic expression proposed by Giaever *et al.* is applicable to numerous cells on a large electrode.[34,35] It assumes that many cells with identical properties form a monolayer with identical intercellular gaps. Changes of the cell-substrate gap and the cell radius can be characterized from the change in measured impedance using the equation of Giaever *et al*. However, discrepancies have been reported between results obtained with the analytic expression and experimental data when size of the electrode can not be approximated as infinity.[36] Urdapilleta *et al.* have proposed an analytic expression that assume a finite electrode size and can be used to analyze the behavior of a population of cells based on the equation of Giaever *et al.*[37] This proposed model assumes that the microscopic response of the cells is propagated by intercellular interactions and is averaged across an electrode. However, the equations proposed by Giaever *et al.* and Urdapilleta *et al.* can only be used when the cultured cells are confluent. Mondal *et al.* have proposed an analytic equation that can be used to estimate the dynamic change of a parameter when a cell on a large electrode adheres and elongates.[38] Mondal's formula provides electrode coverage area of a single cell and cell-substrate gap, and cluster size under a cell culture process. Labayen *et al.* reported the formula against cell monolayer considering effects of size of the electrode and cell.[36] Use of this equation and application of the appropriate boundary conditions for each cell allows estimation of the resistance of the intercellular junctions, the capacitance of the cell membrane, and the cell-substrate gap due to the size of the electrode.

Recently, Shiozawa *et al*. have reported an analytic equation for monitoring a single cell with a microelectrode that is smaller than the cell.[39] The equation of Shiozawa *et al.* enables calculation of single cell parameters such as cell size, cell-substrate gap, and capacitance of the cell. The error rate of the equation based on numerical simulation has been shown to be less than 2.0%.[39] However, the equation was derived on the assumption that the cell was at the center of the electrode. Use of the equation may cause large errors when the electrode is not at the center of a cell.

In this work, we improved the equation of Shiozawa *et al.* and proposed an equation for a single cell on a microelectrode considering the relative positions of the cell and electrode. We used numerical simulations to determine the accuracy of the equation. We found that our proposed equation successfully accounted for the effect of the relative position of the cell and electrode. The equation reduced the error associated with the use of the equation of Shiozawa *et al.* by up to 50%. This work is an extension of our preliminary results reported previously.[40]

This paper is organized as follows. Section II details the major theories that underlie this work and our proposed formula. Section III describes how the effectiveness of the proposed formula was confirmed by numerical simulation. Sections IV and V show simulated and theoretical results and a discussion of the results, respectively. Finally, Section VI describes the conclusions of this study.

## II. Theory

### A. Theory of Giaever *et al.*

The most fundamentally basic analytic equation for the impedance of a cell on a large electrode is given by Giaever *et al.*[34,35] Fig. 1(a), which shows a monolayer of cells on a large electrode, is a schematic diagram of that model. Fig. 1(b) shows the cell-substrate gap under a cell. The origin of the z-axis is at the center of the cell, and the horizontal axis is the radial coordinate $r$. The equation of Giaever *et al.* assumes that disk-shaped cells are distributed continuously as in Fig. 1(a), and the impedance calculated for a single cell in Fig. 1(b) is extended to all cells. The capacitance of the entire cell membrane is equated to that of the serially connected top and bottom cell membranes. A cell adheres to the working electrode (WE) for structural stabilization, and the cell-substrate gap indicates the adhesive strength.[41] A counter electrode (CE) is assumed to be present above the cell monolayer, and the impedance between WE and CE is calculated. The impedance estimated by Giaever *et al.* ($Z_G$) is as follows:

$$Z_{\mathrm{G}}^{-1} = \frac{1}{Z_{\mathrm{n}}}\left[\frac{Z_{\mathrm{n}}}{Z_{\mathrm{n}} + Z_{\mathrm{m}}} + \frac{Z_{\mathrm{m}}}{Z_{\mathrm{n}} + Z_{\mathrm{m}}}\frac{1}{\frac{\gamma r_{\mathrm{c}}}{2}\frac{I_0(\gamma r_{\mathrm{c}})}{I_1(\gamma r_{\mathrm{c}})} + R_{\mathrm{b}}\left(\frac{1}{Z_{\mathrm{n}}} + \frac{1}{Z_{\mathrm{m}}}\right)}\right], \tag{1}$$

where $Z_{\mathrm{m}}$ $[\Omega \cdot \mathrm{m}^2]$ and $Z_{\mathrm{n}}$ $[\Omega \cdot \mathrm{m}^2]$ are the specific impedance of the cell membrane and the electrode, respectively, and $R_{\mathrm{b}}$ $[\Omega \cdot \mathrm{m}^2]$ represents the resistance between cells per unit area. The $I_0(x)$ and $I_1(x)$ are the modified Bessel function of the first kind of order 0 and 1, respectively. The $\gamma$ is a constant, and $r_{\mathrm{c}}$ is the cell radius.

### B. Theory of Shiozawa *et al.*

The equation of Shiozawa *et al.* describes the impedance for a single isolated cell when the working electrode, which is smaller than the cell, is located at the center of the cell, as shown in Fig. 1(c).[39] Here, we summarize its derivation to help readers understand the development of our formulation in subsection IIC. In calculating the impedance, the dependence on the $z$-axis in the cell-substrate gap is ignored because the cell-substrate gap is much smaller than the width of the cell and electrode. The Cylindrical coordinates are used due to take advantage of the rotational symmetry. Fig. 1(d) is a schematic diagram of the relationship between electric potential $V$ and current $I$ in the model. The cell-substrate gap is first divided into regions A and B, and the differential equations for the electric potential and current are solved in each region. The differential equations in region A are expressed as follows

$$-\mathrm{d}V_{\mathrm{A}} = \frac{I_{\mathrm{A}}(r)}{2\pi r h \sigma_{\mathrm{sol}}}\mathrm{d}r, \tag{2}$$

$$V_{\mathrm{e}} - V_{\mathrm{A}}(r) = \frac{Z_{\mathrm{n}}}{2\pi r \mathrm{d}r}\mathrm{d}I_{\mathrm{e}}, \tag{3}$$

$$V_{\mathrm{A}}(r) - V_{\mathrm{c}} = \frac{Z_{\mathrm{m}}}{2\pi r \mathrm{d}r}\mathrm{d}I_{\mathrm{c}}, \tag{4}$$

$$\mathrm{d}I_{\mathrm{A}} = \mathrm{d}I_{\mathrm{e}} - \mathrm{d}I_{\mathrm{c}}, \tag{5}$$

where $V_{\mathrm{A}}(r)$ [V] and $I_{\mathrm{A}}(r)$ [A] are the electric potential and current at the radial position $r$ in region A, respectively. The $V_{\mathrm{e}}$ [V] and $V_{\mathrm{c}}$ [V] are the electric potentials of the working electrode (WE) and electrolyte above the cell, respectively. The $I_{\mathrm{e}}(r)$ [A] and $I_{\mathrm{c}}(r)$ [A] are the current flowing out of the WE and through the cell membrane at $r$, respectively. The $h$ [m] indicates the cell-substrate gap, and $\sigma_{\mathrm{sol}}$ [S/m] is the conductivity of the electrolyte. The $Z_{\mathrm{n}}$ [$\Omega \cdot \mathrm{m}^2$] is the specific impedance of the electrode and is described by the following equation,

$$Z_{\mathrm{n}} = \frac{1}{j\omega C_{\mathrm{dl}}}, \tag{6}$$

where $j$ and $\omega$ [rad/s] are the imaginary unit and angular frequency of the AC stimulation, respectively, and $C_{\mathrm{dl}}$ [$\mathrm{F/m^2}$] is the electrical double layer capacitance per unit area, given by

$$C_{\mathrm{dl}} = \frac{\varepsilon_0 \varepsilon_{\mathrm{sol}}}{\lambda_D}, \tag{7}$$

where $\varepsilon_0$ [F/m], $\varepsilon_{\mathrm{sol}}$, and $\lambda_0$ [m] are the permittivity of the vacuum, the relative permittivity of the electrolyte, and the Debye length, respectively. The $Z_{\mathrm{m}}$ [$\Omega \cdot \mathrm{m}^2$] indicates the specific impedance of the cell membrane and is described by the following equation,

$$Z_{\mathrm{m}} = \frac{1}{j\omega C_{\mathrm{m}}}, \tag{8}$$

where $C_{\mathrm{m}}$ $[\mathrm{F/m^2}]$, the capacitance of the cell membrane per unit area, is given by

$$C_{\mathrm{m}} = \left(\frac{1}{C_{\mathrm{m,upper}}} + \frac{1}{C_{\mathrm{m,botom}}}\right)^{-1}, \tag{9}$$

where $C_{\mathrm{m,upper}}$ $[\mathrm{F/m^2}]$ and $C_{\mathrm{m,bottom}}$ $[\mathrm{F/m^2}]$ are the capacitance of the upper and bottom cell membranes, respectively. The capacitance of cell membrane can often be approximated as $C_{\mathrm{m,upper}} \gg C_{\mathrm{m,bottom}}$ in a cell having a bulging shape. The general solution to Eqs. (2) ~ (5) for $V_{\mathrm{A}}(r)$ is given by

$$V_{\mathrm{A}}(r) = C_{\mathrm{A}} I_0(\gamma_{\mathrm{A}} r) + \frac{Z_{\mathrm{n}} V_{\mathrm{c}} + Z_{\mathrm{m}} V_{\mathrm{e}}}{Z_{\mathrm{n}} + Z_{\mathrm{m}}}, \tag{10}$$

where $C_{\mathrm{A}}$ is a constant of integration, and $I_{\mathrm{n}}(x)$ is the modified Bessel function of the first kind of order $n$. The $\gamma_{\mathrm{A}}$ is defined as

$$\gamma_{\mathrm{A}} = \sqrt{\frac{1}{h\sigma_{\mathrm{sol}}}\left(\frac{1}{Z_{\mathrm{n}}} + \frac{1}{Z_{\mathrm{m}}}\right)}. \tag{11}$$

Similarly, the general solution to the equations for $V_{\mathrm{B}}(r)$ is given by

$$V_{\mathrm{B}}(r) = C_{\mathrm{B}} I_0(\gamma_{\mathrm{B}} r) + D_{\mathrm{B}} K_0(\gamma_{\mathrm{B}} r) + V_{\mathrm{c}}, \tag{12}$$

where $V_{\mathrm{B}}(r)$ is the potential at the radial position $r$ in region B, and $K_{\mathrm{n}}(x)$ is the modified Bessel function of the second kind of order $n$. Here, $C_{\mathrm{B}}$ and $D_{\mathrm{B}}$ are constants of integration, and $\gamma_{\mathrm{B}}$ is defined as

$$\gamma_{\mathrm{B}} = \sqrt{\frac{1}{h\sigma_{\mathrm{sol}} Z_{\mathrm{m}}}}. \tag{13}$$

At the interface between regions A and B, the boundary conditions of the electric potential and current are given by

$$V_{\mathrm{A}}(r_{\mathrm{e}}) = V_{\mathrm{B}}(r_{\mathrm{e}}), \tag{14}$$

$$I_{\mathrm{A}}(r_{\mathrm{e}}) = I_{\mathrm{B}}(r_{\mathrm{e}}), \tag{15}$$

where $I_{\mathrm{B}}(r)$ is the current at the radial position $r$ in region B, and $r_{\mathrm{e}}$ [m] is the electrode radius. We used the following boundary condition at the edge of the cell.

$$V_{\mathrm{B}}(r_{\mathrm{c}}) = V_{\mathrm{a}}, \tag{16}$$

where $r_{\mathrm{c}}$ is cell radius, and $V_{\mathrm{a}}$ is the potential at the edge of the cell. Assuming $V_{\mathrm{a}} = V_{\mathrm{c}}$ for simplicity, the cell-substrate impedance $Z_{\mathrm{c}}$ is calculated using

$$Z_{\mathrm{c}} = \frac{V_{\mathrm{e}} - V_{\mathrm{c}}}{I_{\mathrm{e}}}. \tag{17}$$

Here the $I_{\mathrm{e}}$ is obtained from Eq. (3) as follows:

$$I_{\mathrm{e}} = 2\pi \int_0^{r_{\mathrm{e}}} \frac{V_{\mathrm{e}} - V_{\mathrm{A}}(r)}{Z_{\mathrm{n}}} r\mathrm{d}r. \tag{18}$$

Finally, the impedance of Shiozawa *et al.*, $Z_{\mathrm{c}}$ [Ω], the cell-substrate impedance using a microelectrode, is given by,

$$Z_{\mathrm{c}}^{-1} = -\frac{2\pi r_{\mathrm{e}}}{Z_{\mathrm{n}}\gamma_{\mathrm{A}}}\frac{1}{\Delta}k(1-k^2)[I_1(\gamma_{\mathrm{B}}r_{\mathrm{e}})K_0(\gamma_{\mathrm{B}}r_{\mathrm{c}}) + I_0(\gamma_{\mathrm{B}}r_{\mathrm{c}})K_1(\gamma_{\mathrm{B}}r_{\mathrm{e}})]I_1(\gamma_{\mathrm{A}}r_{\mathrm{e}}) + \frac{\pi r_{\mathrm{e}}^2}{(Z_{\mathrm{n}} + Z_{\mathrm{m}})}, \tag{19}$$

where $k$ is a constant defined as

$$k = \frac{\gamma_{\mathrm{B}}}{\gamma_{\mathrm{A}}} = \sqrt{\frac{Z_{\mathrm{n}}}{Z_{\mathrm{n}} + Z_{\mathrm{m}}}} = \sqrt{\frac{C_{\mathrm{m}}}{C_{\mathrm{dl}} + C_{\mathrm{m}}}}, \tag{20}$$

and $\Delta$ is given by

$$\Delta = I_1(\gamma_{\mathrm{A}}r_{\mathrm{e}})[I_0(\gamma_{\mathrm{B}}r_{\mathrm{e}})K_0(\gamma_{\mathrm{B}}r_{\mathrm{c}}) - I_0(\gamma_{\mathrm{B}}r_{\mathrm{c}})K_0(\gamma_{\mathrm{B}}r_{\mathrm{e}})] - kI_0(\gamma_{\mathrm{A}}r_{\mathrm{e}})[I_1(\gamma_{\mathrm{B}}r_{\mathrm{e}})K_0(\gamma_{\mathrm{B}}r_{\mathrm{c}}) + I_0(\gamma_{\mathrm{B}}r_{\mathrm{c}})K_1(\gamma_{\mathrm{B}}r_{\mathrm{e}})]. \tag{21}$$

## C. Proposed equation to consider dependence on cell position

The equation of Shiozawa *et al.* is valid only when the WE is at the center of a cell, but that condition is not always satisfied in actual experiment. In this section, we propose an impedance equation that is valid even when the location of the WE is not the center of the cell. Fig. 2(a) shows a schematic diagram of the situation where the center of the cell and electrode are displaced along the $x$-axis. We must now use Cartesian coordinates because there is no longer rotational symmetry if the electrode is not at the center of the cell. Because the cell and WE are symmetric about the $x$-axis, we can concentrate our attention on the upper half of the geometry, as shown in Fig. 2(b). We now calculate the impedance for this geometry. First, we consider an infinitesimally small slice of the area between the electrode and the edge of the cell, as shown in Fig. 2(c), where $\mathrm{d}\theta$ indicates the infinitesimally small angle of the slice. Second, we assume that the path of the electric current within this slice is linear in the radial direction. Now, remember that the equation of Shiozawa *et al.* describes the impedance when the electrode is at the center of the cell, as in Fig. 1(c). In such a situation, the electric current flows in a strictly radial direction from the center of the WE to the edge of the cell. Therefore, the equation of Shiozawa *et al.* could be used to describe the impedance within the small slice in Fig. 2(c). We therefore assumed that the admittance in the small slice in Fig. 2(c) could be approximated by the admittance of the equation of Shiozawa *et al.* within the angle $\mathrm{d}\theta$. The admittance $\mathrm{d}Y$ within the angle $\mathrm{d}\theta$ is thus given by

$$\mathrm{d}Y = \frac{\mathrm{d}\theta}{2\pi}\frac{1}{Z_\mathrm{c}}, \tag{22}$$

where $Z_\mathrm{c}$ $[\Omega]$ is the cell-substrate impedance of Shiozawa *et al.* given by Eq. (19). The admittance due to the electric current flowing from the center of the WE to the arc CD in

Fig. 2(c) is thus given by the integral of Eq. (22) over $0 \leq \theta \leq \pi$:

$$Y_{\mathrm{CD}} = \int_0^{\pi} \frac{1}{2\pi Z_{\mathrm{c}}\left(r_{\mathrm{e}},\ r_{\mathrm{c}}'(\theta)\right)} \mathrm{d}\theta, \tag{23}$$

where $r_{\mathrm{c}}'(\theta)$ [m] is the distance from point O to B in Fig. 2(c) and is a function of $\theta$. The dependence of $r_{\mathrm{c}}'(\theta)$ on $\theta$ can be explicitly calculated with the help of Fig. 2(d). Application of the cosine relationship to the triangle OAB enables the cell radius $r_{\mathrm{c}}'(\theta)$ to be written as

$$r_{\mathrm{c}}'(\theta) = X_{\mathrm{c}} \cos\theta + \sqrt{X_{\mathrm{c}}^2(\cos^2\theta - 1) + r_{\mathrm{c}}^2}, \tag{24}$$

where $X_{\mathrm{c}}$ [m] is the relative displacement between the cell and electrode. The whole admittance is expressed by parallel connection of $Y_{\mathrm{CD}}$ from the upper and the lower half of the circle. The equation for the impedance $Z_{\mathrm{s}}$ [Ω] is then given:

$$Z_{\mathrm{s}}^{-1} = 2Y_{\mathrm{CD}} = 2\int_0^{\pi} \frac{1}{2\pi Z_{\mathrm{c}}\left(r_{\mathrm{e}},\ r_{\mathrm{c}}'(\theta)\right)} \mathrm{d}\theta. \tag{25}$$

In practice, the integral is evaluated by numerical integration.

## III. Simulation

We used COMSOL Multiphysics 6.0 to perform a numerical simulation to confirm the effectiveness of our proposed equation. We used the same simulation model, boundary conditions, and parameter values used by Shiozawa *et al.*[39] The only difference was that we used Cartesian coordinates to simulate the displacement of the cell relative to the electrode location. Fig. 3 shows the simulation model. The cell radius $r_{\mathrm{c}}$ and cell-substrate gap $h$ were $5.0\ \mu\mathrm{m}$ and $100\ \mathrm{nm}$, respectively. The electrode radius $r_{\mathrm{e}}$ and dislocation of the cell relative to the electrode $X_{\mathrm{c}}$ were variables. An insulation boundary condition was imposed on the bottom of the model shown by the red line in Fig. 3, and we assumed that the electric potential on the boundaries indicated by the blue lines in Fig. 3 was zero volts. This assumption is equivalent to approximating $V_{\mathrm{a}} = V_{\mathrm{c}} = 0$. Table 1 shows the parameters used in our simulation. In this work, we approximated $C_{\mathrm{m}}$ by $C_{\mathrm{m,bottom}}$ because the capacitance of the upper cell membrane is often much larger than that of the bottom cell membrane. Although $C_{\mathrm{m}}$ was represented by Eq. (9), it could be approximated as $C_{\mathrm{m,upper}} \gg C_{\mathrm{m,bottom}}$ because the shape of a cell is hemispherical when it adheres onto the substrate. We therefore performed a simulation with $C_{\mathrm{m}} \approx C_{\mathrm{m,bottom}}$. Even with this assumption, the generality is not lost in confirming the accuracy of our theory. The simulation was performed in the following steps. First, an AC voltage with an amplitude of $5.0\ \mathrm{mV}$ and zero DC bias was applied to WE $(V_{\mathrm{e}})$ in the frequency range of $10^2\ \mathrm{Hz}$ to $10^7\ \mathrm{Hz}$. We then simulated the time dependence of the electric current flowing on the WE. Finally, we obtained the frequency characteristics of the impedance by calculating the response signal using a discrete Fourier transform. The $X_{\mathrm{c}}$ was swept from $0.0\ \mu\mathrm{m}$ to the position where the edge of the WE touched the edge of the cell, and the $r_{\mathrm{e}}$ was swept from $1.0\ \mu\mathrm{m}$ to $4.0\ \mu\mathrm{m}$ in steps of $1.0\ \mu\mathrm{m}$.

## IV. Results

Fig. 4 shows Bode plots of the theoretical impedance calculated from our proposed equation (solid lines) and the simulated impedance (dots) for electrode radii of $r_{\mathrm{e}} = 1.0\ \mu\mathrm{m}$ (Fig. 4 (a)), $r_{\mathrm{e}} = 2.0\ \mu\mathrm{m}$ (Fig. 4 (b)), $r_{\mathrm{e}} = 3.0\ \mu\mathrm{m}$ (Fig. 4 (c)), and $r_{\mathrm{e}} = 4.0\ \mu\mathrm{m}$ (Fig. 4 (d)). The red solid lines correspond to the results from the equation of Shiozawa *et al.*, which is a special case of our equation with $X_{\mathrm{c}} = 0.0\ \mu\mathrm{m}$. In Fig. 4, the equation of Shiozawa *et al.* agrees well with the numerical simulation only when $X_{\mathrm{c}} = 0.0\ \mu\mathrm{m}$. As the displacement of the cell relative to the electrode location $X_{\mathrm{c}}$ increases, the discrepancy between the equation of Shiozawa *et al.* and the numerical simulation increases. Our proposed equation, however, agreed very well with the simulated results, even for $X_{\mathrm{c}} > 0$. To confirm the effectiveness of our equation, we used the following equation to calculate the error rate between the simulated and theoretical impedance.

$$\text{Error rate [\%]} = \frac{|Z_{\mathrm{simulation}}| - |Z_{\mathrm{theory}}|}{|Z_{\mathrm{theory}}|} \times 100, \tag{26}$$

where $Z_{\mathrm{simulation}}\ [\Omega]$ and $Z_{\mathrm{theory}}\ [\Omega]$ indicate the simulated and theoretical impedance, respectively. Fig. 5(a), (c), (e), and (g) show the error rates of the equation of Shiozawa *et al.*, and Fig. 5(b), (d), (f), and (h) indicate those of our proposed equation. The electrode radii used were $r_{\mathrm{e}} = 1.0\ \mu\mathrm{m}$ (Fig. 4(a) and (b)), $r_{\mathrm{e}} = 2.0\ \mu\mathrm{m}$ (Fig. 4(c) and (d)), $r_{\mathrm{e}} = 3.0\ \mu\mathrm{m}$ (Fig. 4(e) and (f)), and $r_{\mathrm{e}} = 4.0\ \mu\mathrm{m}$ (Fig. 4(g) and (h)). The obvious reduction of the error rate by our proposed equation versus the equation of Shiozawa *et al.* demonstrated the effectiveness of our proposed equation Eq. (25)

## V. Discussion

To clearly explain the frequency characteristics of the impedance, we now discuss the changes in the simulation results shown in Fig. 4(a). In Fig. 4(a), the impedance of the electric double layer was dominant from $10^2$ Hz to $10^4$ Hz. Although the impedance of an electric double layer at an electrode surface is a function of the AC frequency and depends on electrode radius, it is insensitive to positional changes of the cell. The resistance of the cell-substrate gap was dominant from $10^4$ Hz to $10^6$ Hz (Fig. 4(a)). It is apparent that the resistance decreased gradually because of the changes of the position of the cell. Fig. 6(a) and (b) show the electric current densities at the cell-substrate gap at a fixed $r_e$ of 1.0 μm and $f = 10^5$ Hz for (a) $X_c = 1.0$ μm and (b) $X_c = 4.0$ μm. It is apparent that the electric current density was more concentrated in $x < 0$ due to changes of the position of the cell. The resistance decreased because the electric current concentrated on an area where the distance between edge of the cell and edge of the electrode decreased. In Fig. 4(a), the impedance of the cell membrane was dominant from $10^6$ Hz to $10^7$ Hz because slopes of the impedance are close to $-1$ for most of $X_c$. However, the impedance of the cell membrane is less dominant for $X_c = 4.0$ μm, where the edge of the cell overlapped with the edge of the electrode. The explanation for this behavior is that an electric current leaked at rate $R$ from the side of the gap without penetrating the cell membrane $(I_E)$ and flowed through the cell membrane $(I_M)$. The magnitude of $R$ is given by

$$R = \frac{|I_M|}{|I_M| + |I_E|}, \tag{27}$$

where $|I_M|$ and $|I_E|$ are averages over a periodic sinusoidal change in time. Fig. 7 shows values of $R$ for various values of $X_c$ at a fixed $r_e = 1.0$ μm and $f = 10^7$ Hz.

In Fig. 7, $|I_{\mathrm{M}}|$ was dominant when $X_{\mathrm{c}} = 0.0$– $3.0$ μm because the values of $R$ were high ($0.97$– $0.86$) within that range of $X_{\mathrm{c}}$. Meanwhile, $|I_{\mathrm{E}}|$ occurred in parallel to $|I_{\mathrm{M}}|$ when $X_{\mathrm{c}} = 4.0$ μm because $R$ decreased to 0.50. At ~$10^7$ Hz, the impedance was therefore determined by the combined resistance of the impedance of the cell membrane and the solution resistance in parallel.

At $\sim 10^5$ Hz, an error of as much as 20% is seen in Fig. 5(b), and a similar error also appears in Fig. 5(d), (f), (h). Fig. 8(a) and (b) show the paths of the electric current in the cell-substrate gap at a fixed $r_{\mathrm{e}}$ of 1.0 μm and $f = 10^5$ Hz for (a) $X_{\mathrm{c}} = 1.0$ μm and (b) $X_{\mathrm{c}} = 4.0$ μm. It is apparent that the electric current path formed curved lines by increasing the displacement of the cell relative to the location of the electrode. This curvature indicates that the paths of the current deviated from our assumed radial direction when the displacement of the cell relative to the location of the electrode increased. However, this error was unavoidable because the equation was derived by assuming the path of the electric current was radial. The same error is apparent in Fig. 5(b), (d), (f), (h) for $X_{\mathrm{c}} = 4.0$ μm at $\sim 10^6$ Hz. The electric current leaks from the edge of the cell under these conditions (vide supra), and its path becomes curved, as shown in Fig. 8(b).

## VI. Conclusions

To describe cell-substrate impedance, we proposed an equation that was a function of the position of the cell. We used a simulation to verify the effectiveness of the proposed equation. The simulation showed that the response of the calculated impedance to changes in the position of a cell was qualitatively correct. Thus, our proposed equation largely reduced the error compared to the previously proposed Shiozawa's equation. Even with our improved impedance equation, however, slight errors remain when the path of the electric current was not linear in radial direction, which is fundamentally unavoidable. This pattern was a result of our assumption that the path of the electric current was linear. Our analytic equation can thus be used in more realistic situations, and the characteristic parameters of a single cell may be deduced from experimental data with the help of our impedance equation.

## AUTHOR DECLARATIONS

### Conflict of Interest

The authors have no conflicts to disclose.

### Author Contributions

Y. Sugahara and S. Uno contributed equally to this work.

## DATA AVAILABILITY

The data that support the finding of this study are available from the corresponding author upon reasonable request.

## TABLE Caption

**TABLE. 1.** Parameters used in simulation

## FIGURE Captions

**FIG. 1.** (a) Schematic diagram of the model of Giaever *et al.*[34] WE is a working electrode. (b) Schematic diagram of the cell-substrate gap in the model of Giaever *et al*. The red arrow shows the electric current path. (c) Schematic diagram of the model of Shiozawa *et al.*[39] Red arrow shows the electric current path. (d) Relationship between the electric potential and current in the model of Shiozawa *et al*. The symbols are defined in Section IIB.

**FIG. 2.** (a) Relationship between the position of a single cell and an electrode. (b) Simplified relationship between the position of a single cell and an electrode. (c) Schematic diagram of a thin slice between the cell and substrate viewed from the top. (d) Schematic diagram of the cell radius $r_{\mathrm{c}}'(\theta)$ with the center of the working electrode (WE) as the starting point. $r_{\mathrm{c}}$ is the cell radius starting from A, and $X_{\mathrm{c}}$ is the relative distance between the cell and an electrode along the $x$-axis.

**FIG. 3.** Simulation model of the cell-substrate gap.

**FIG. 4.** Bode plots of the theoretical impedance (line plot) calculated from Eq. (25) and simulated impedance (dot plot) at the radius of each working electrode (WE). The impedance is plotted at a WE radius $r_{\mathrm{e}}$ of (a) 1.0 μm, (b) 2.0 μm, (c) 3.0 μm, and (d) 4.0 μm. Legends indicate the relative distance between a cell and an electrode along $x$-axis. Titles show the WE radius. When $X_c = 0.0$ μm, our equation is equivalent to the equation of Shiozawa *et al.*

**FIG. 5.** Error rates between the simulated and theoretical impedance. Error rates of the equation of Shiozawa *et al.* are plotted at a working electrode (WE) radius $r_{\mathrm{e}}$ of (a) 1.0 μm, (c) 2.0 μm, (e) 3.0 μm, and (g) 4.0 μm, which error rates of our proposed equation Eq. (25) are plotted at a WE radius $r_{\mathrm{e}}$ of (b) 1.0 μm, (d) 2.0 μm, (f) 3.0 μm, and (h) 4.0 μm. Legends indicate the relative distance between a cell and an electrode along the $x$-axis.

**FIG. 6.** (a) Color map of the electric current density in the cell substrate gap when $r_{\mathrm{e}} = 1.0$ μm, $X_{\mathrm{c}} = 1.0$ μm, and $f = 10^5$ Hz. (b) Electric current density in the cell substrate gap when $r_{\mathrm{e}} = 1.0$ μm, $X_{\mathrm{c}} = 3.0$ μm, and $f = 10^5$ Hz.

**FIG. 7.** The rate of electric current $R$ in Eq. (27).

**FIG. 8.** (a) Electric current path in the cell substrate gap when $r_{\mathrm{e}} = 1.0$ μm, $X_{\mathrm{c}} = 1.0$ μm, $f = 10^5$ Hz. (b) Electric current path in the cell substrate gap when $r_{\mathrm{e}} = 1.0$ μm, $X_{\mathrm{c}} = 4.0$ μm, $f = 10^5$ Hz.

**TABLE. 1.**

| Parameter | Value |
|---|---|
| Double layer capacitance per unit area $[\mathrm{F/m^2}]$ | 0.89 |
| Solution relative permittivity | 78 |
| Solution conductivity $[\mathrm{S/m}]$ | 1.5 |
| Cell membrane thickness $[\mathrm{nm}]$ | 5.0 |
| Cell membrane relative permittivity | 5.0 |
| Cell membrane conductivity $[\mathrm{S/m}]$ | $1.0 \times 10^{-8}$ |

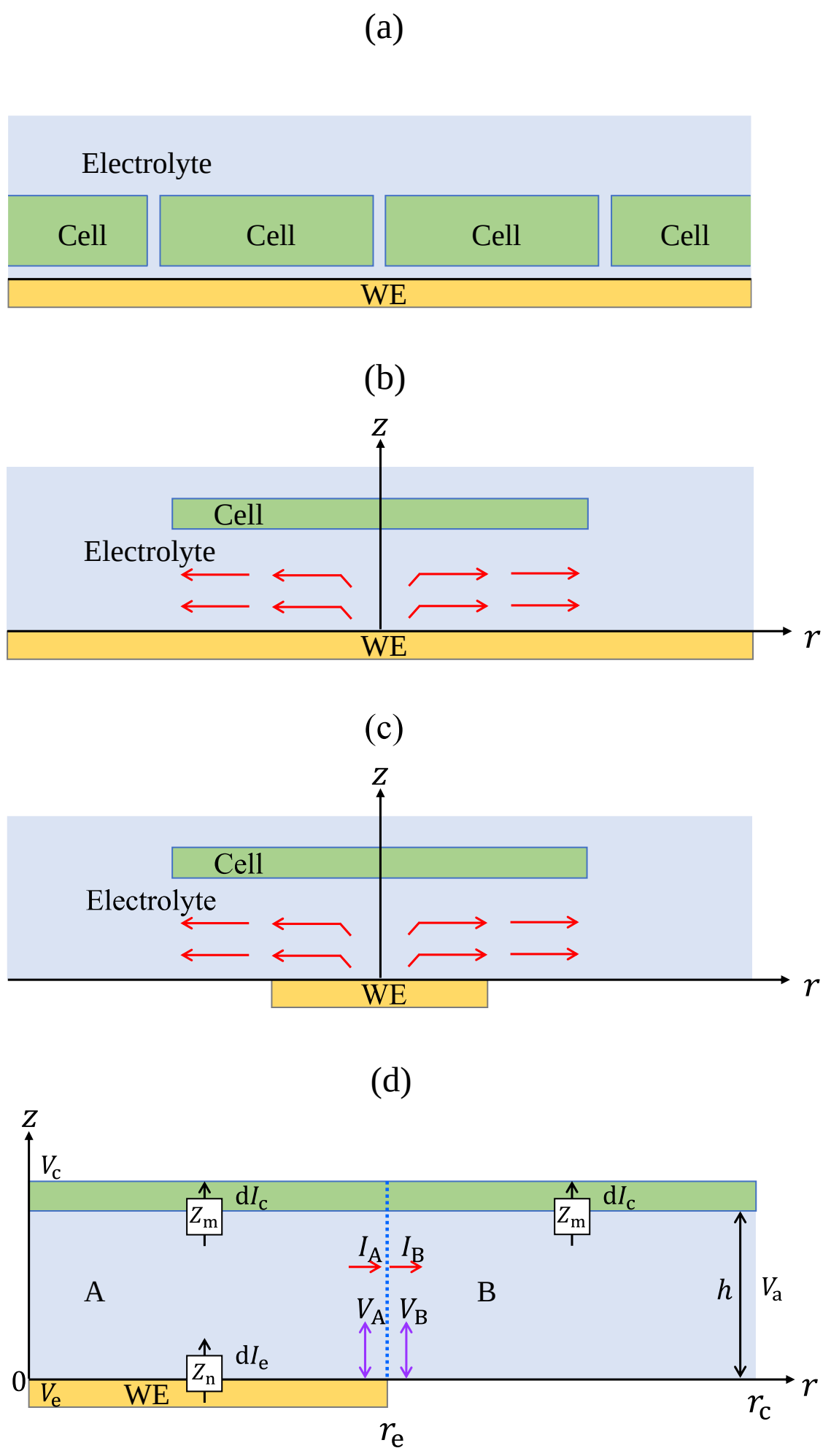

**FIG. 1.**

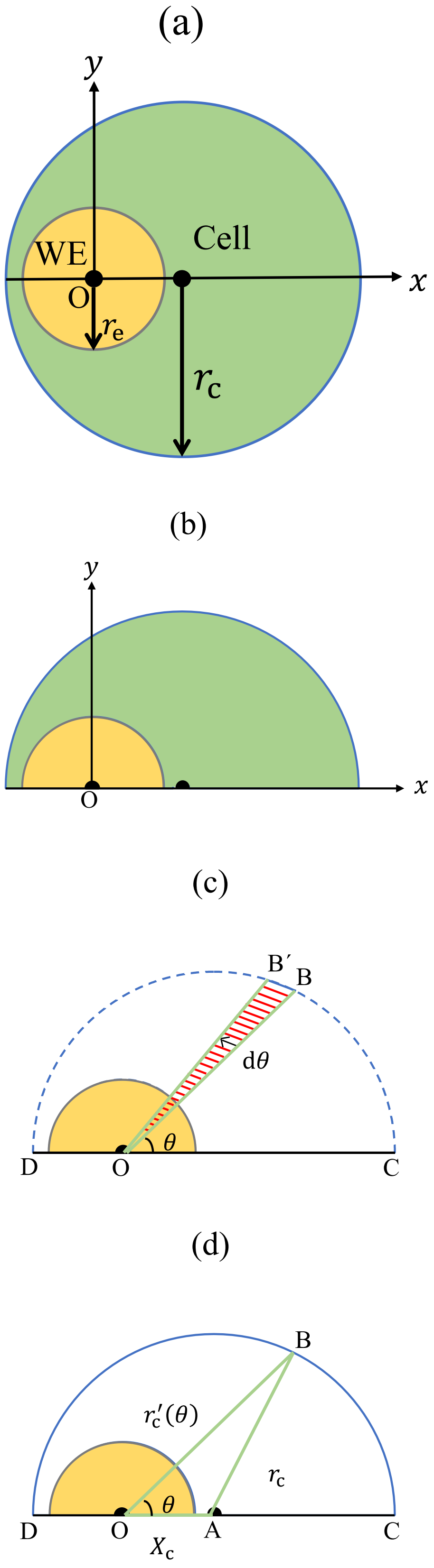

**FIG. 2.**

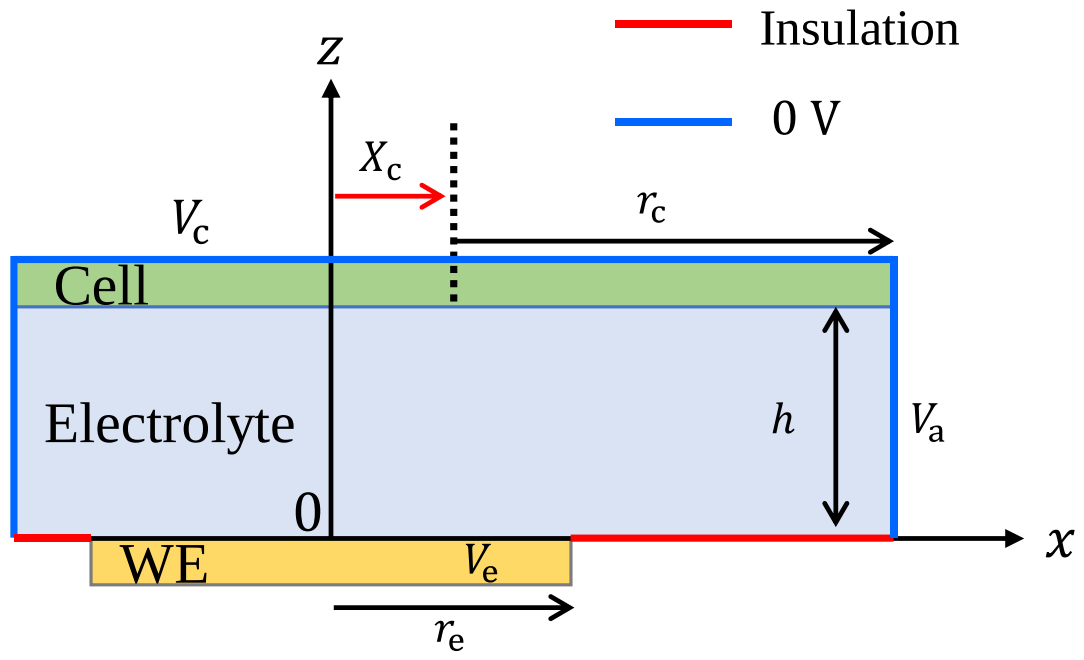

**FIG. 3.**

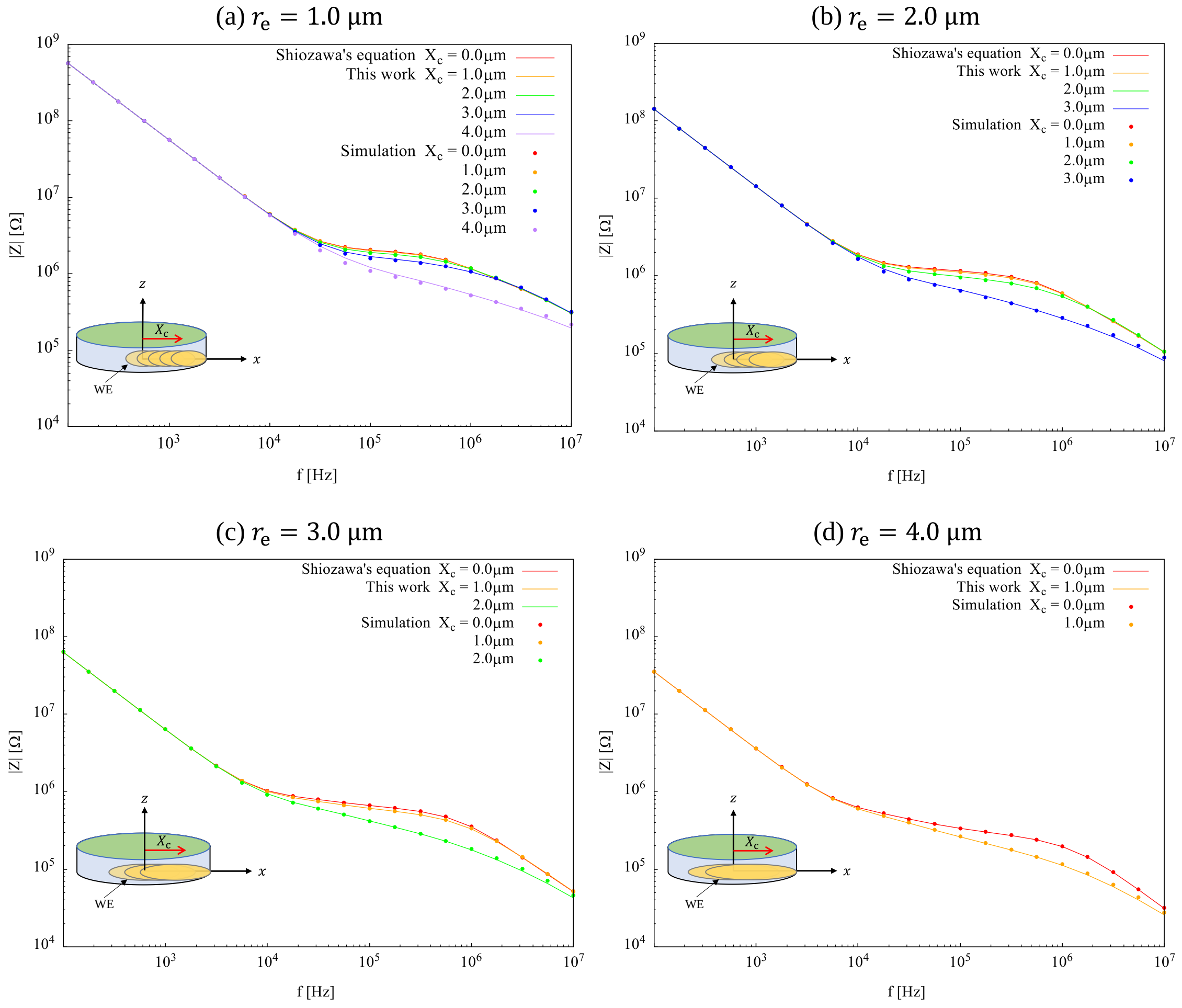

(a) $r_e$ = 1.0 μm
Shiozawa's equation $X_c$ = 0.0μm
This work $X_c$ = 1.0μm
2.0μm
3.0μm
4.0μm
Simulation $X_c$ = 0.0μm
1.0μm
2.0μm
3.0μm
4.0μm
|Z| [Ω]
f [Hz]
z
x
$X_c$
WE
(b) $r_e$ = 2.0 μm
Shiozawa's equation $X_c$ = 0.0μm
This work $X_c$ = 1.0μm
2.0μm
3.0μm
Simulation $X_c$ = 0.0μm
1.0μm
2.0μm
3.0μm
|Z| [Ω]
f [Hz]
(c) $r_e$ = 3.0 μm
Shiozawa's equation $X_c$ = 0.0μm
This work $X_c$ = 1.0μm
2.0μm
Simulation $X_c$ = 0.0μm
1.0μm
2.0μm
|Z| [Ω]
f [Hz]
(d) $r_e$ = 4.0 μm
Shiozawa's equation $X_c$ = 0.0μm
This work $X_c$ = 1.0μm
Simulation $X_c$ = 0.0μm
1.0μm
|Z| [Ω]
f [Hz]

**FIG. 4.**

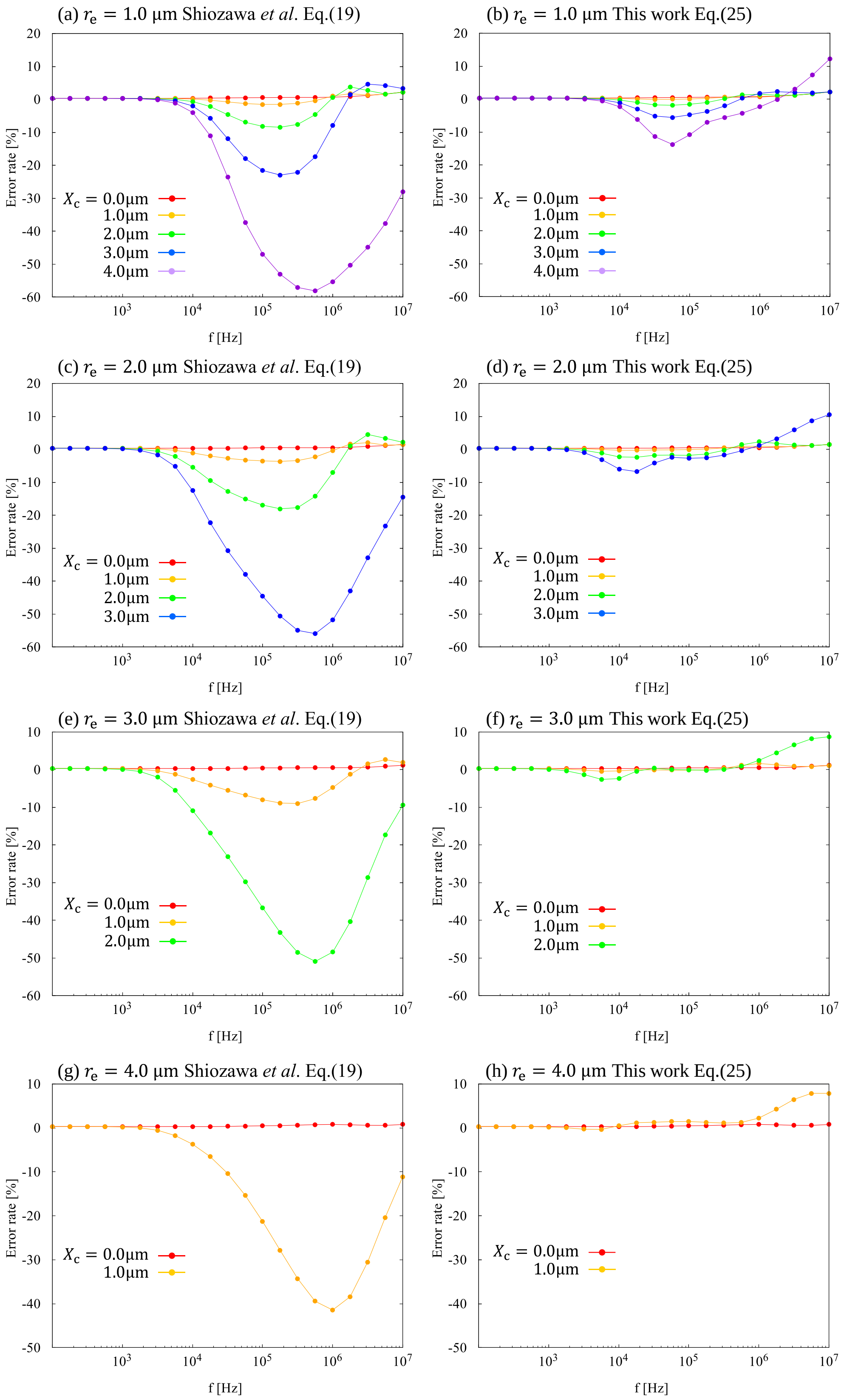
(a) $r_e$ = 1.0 μm Shiozawa *et al.* Eq.(19)
(b) $r_e$ = 1.0 μm This work Eq.(25)
(c) $r_e$ = 2.0 μm Shiozawa *et al.* Eq.(19)
(d) $r_e$ = 2.0 μm This work Eq.(25)
(e) $r_e$ = 3.0 μm Shiozawa *et al.* Eq.(19)
(f) $r_e$ = 3.0 μm This work Eq.(25)
(g) $r_e$ = 4.0 μm Shiozawa *et al.* Eq.(19)
(h) $r_e$ = 4.0 μm This work Eq.(25)
Error rate [%]
f [Hz]
$X_c$ = 0.0μm
1.0μm
2.0μm
3.0μm
4.0μm

**FIG. 5**

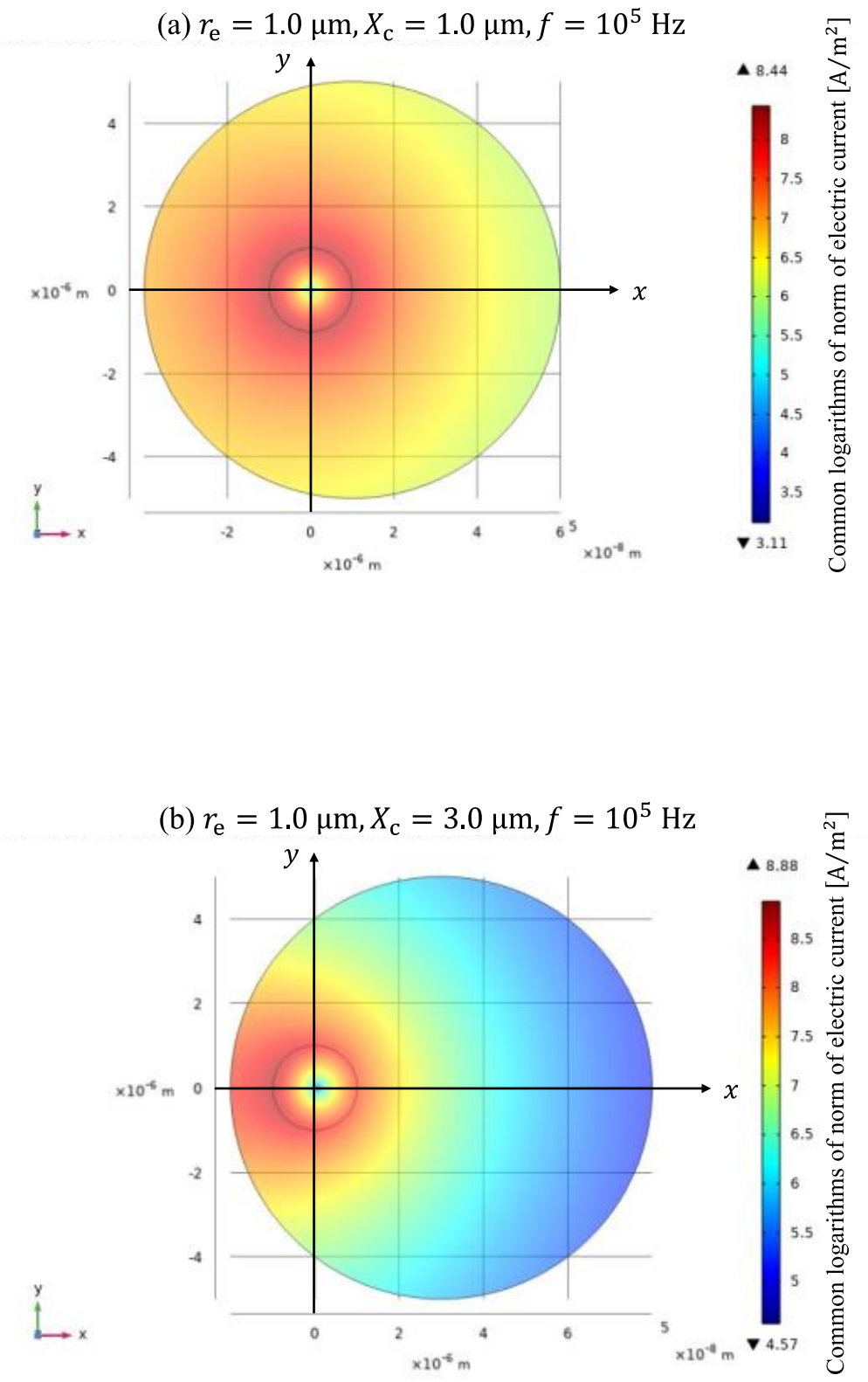
(a) $r_e = 1.0$ μm, $X_c = 1.0$ μm, $f = 10^5$ Hz
y
x
×10⁻⁶ m
Common logarithms of norm of electric current [A/m²]
▲ 8.44
▼ 3.11
(b) $r_e = 1.0$ μm, $X_c = 3.0$ μm, $f = 10^5$ Hz
y
x
×10⁻⁶ m
Common logarithms of norm of electric current [A/m²]
▲ 8.88
▼ 4.57

**FIG. 6.**

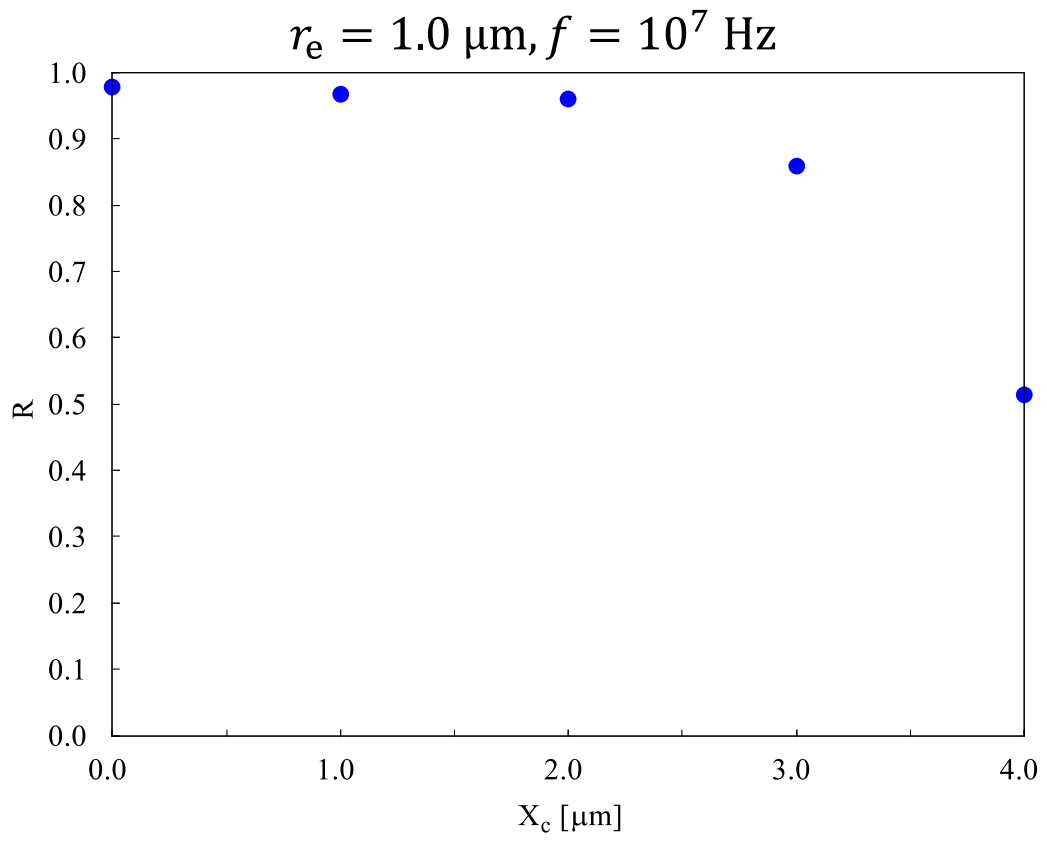
$r_{\mathrm{e}} = 1.0$ μm, $f = 10^7$ Hz
1.0
0.9
0.8
0.7
0.6
R 0.5
0.4
0.3
0.2
0.1
0.0
0.0
1.0
2.0
3.0
4.0
$X_c$ [μm]

**FIG. 7.**

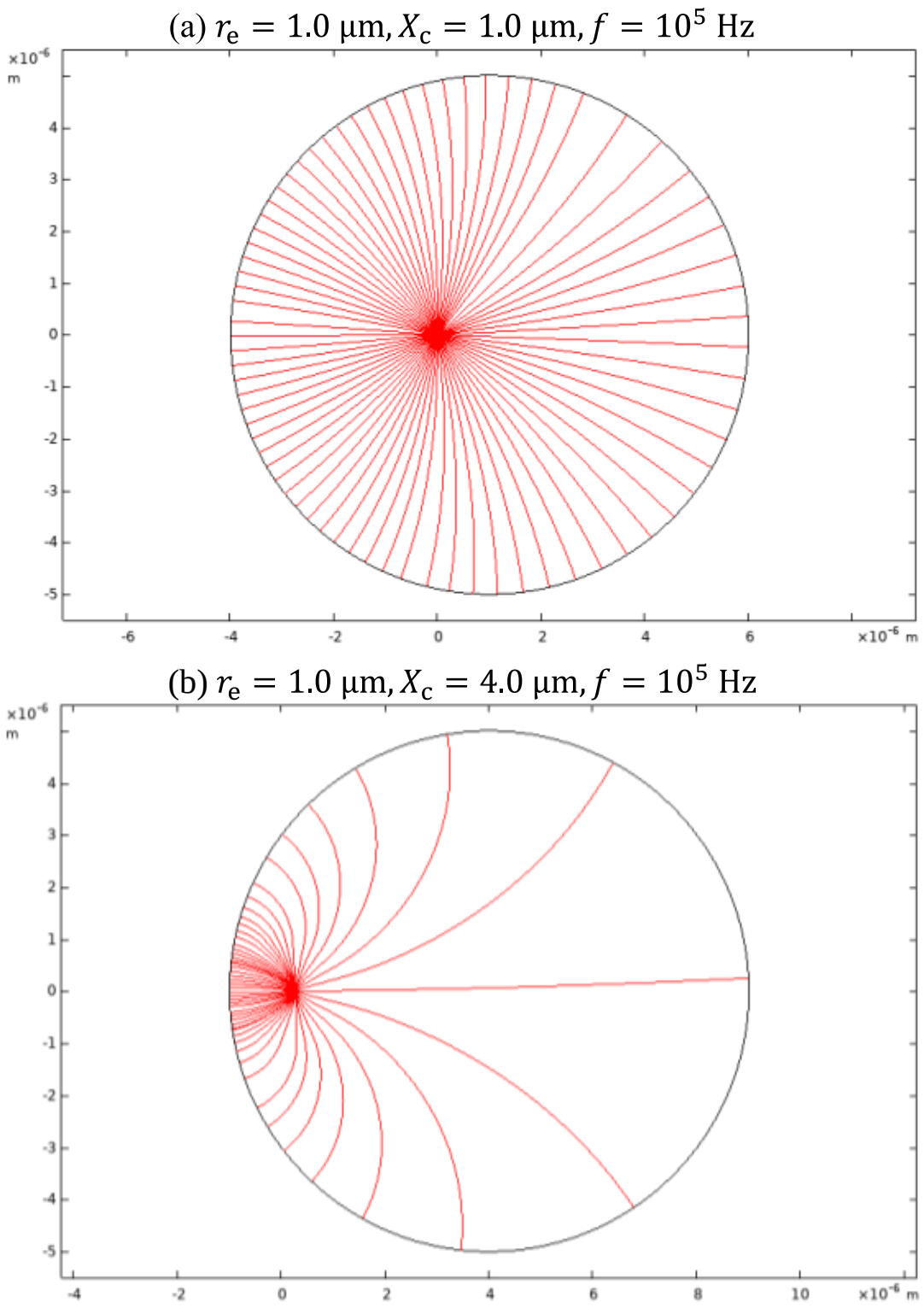

**FIG. 8.**